\documentstyle[12pt]{article}

\parskip=\medskipamount                 % space between paragraphs (LaTeX)
\def\ket#1{| #1\rangle}                 % | >
\def\7#1#2{\mathop{\null#2}\limits^{#1}}        % puts #1 atop #2
\def\greaterthansquiggle{\raise.3ex\hbox{$>$\kern-.75em\lower1ex\hbox{$\sim$}}}
\def\lessthansquiggle{\raise.3ex\hbox{$<$\kern-.75em\lower1ex\hbox{$\sim$}}}
\def\Nsquiggle{\raise.3ex\hbox{$N$\kern-.75em\lower1ex\hbox{$\,\sim$}}}

\def\bo{{\raise.15ex\hbox{\large$\Box$}}}

\def\TH{{\raise.2ex\hbox{$\displaystyle \bigodot$}\mskip-4.7mu \llap H \;}}
\def\face{{\raise.2ex\hbox{$\displaystyle \bigodot$}\mskip-2.2mu \llap {$\ddot
        \smile$}}}

\def\Tilde#1{\widetilde {#1}}
\def\Hat#1{\rlap{\kern.10em$\widehat{\phantom G}$}#1}
\def\HAt#1{\rlap{\kern.05em$\widehat{\phantom G}$}#1}

\def\cap#1{\rlap{\kern.1em$\widehat{\phantom{G\vrule height.8em}}$}#1{}}
\def\Cap#1{\rlap{\kern.05em$\widehat{\phantom{G\vrule height.8em}}$}#1{}}
\def\bra#1{\left\langle #1\right|}
\def\ket#1{\left| #1\right\rangle}

\def\VEV#1{\left\langle #1\right\rangle}

\def\leftrightarrowfill{$\mathsurround=0pt \mathord\leftarrow \mkern-6mu
        \cleaders\hbox{$\mkern-2mu \mathord- \mkern-2mu$}\hfill
        \mkern-6mu \mathord\rightarrow$}
\def\overleftrightarrow#1{\vbox{\ialign{##\crcr
        \leftrightarrowfill\crcr\noalign{\kern-1pt\nointerlineskip}
        $\hfil\displaystyle{#1}\hfil$\crcr}}}

\def\frac#1#2{{\textstyle{#1\over\vphantom2\smash{\raise.20ex
        \hbox{$\scriptstyle{#2}$}}}}}

\def\sfrac#1#2{{\vphantom1\smash{\lower.5ex\hbox{\small$#1$}}\over
        \vphantom1\smash{\raise.4ex\hbox{\small$#2$}}}}
\def\bfrac#1#2{{\vphantom1\smash{\lower.5ex\hbox{$#1$}}\over
        \vphantom1\smash{\raise.3ex\hbox{$#2$}}}}
\def\afrac#1#2{{\vphantom1\smash{\lower.5ex\hbox{$#1$}}\over#2}}

\catcode`@=11
\def\underline#1{\relax\ifmmode\@@underline#1\else
        $\@@underline{\hbox{#1}}$\relax\fi}
\catcode`@=12

\def\nis{\nointerlineskip}
\def\Abar{\vbox{\nis\moveright.33em\vbox{
        \hrule width.35em height.04em}\nis\kern.05em\hbox{$A$}}{}}
\def\Dbar{\vbox{\nis\moveright.20em\vbox{
        \hrule width.50em height.04em}\nis\kern.05em\hbox{$D$}}{}}
\def\Gbar{\vbox{\nis\moveright.20em\vbox{
        \hrule width.50em height.04em}\nis\kern.05em\hbox{$G$}}{}}
\def\mbar{\vbox{\nis\moveright.15em\vbox{
        \hrule width.60em height.04em}\nis\kern.05em\hbox{$m$}}{}}
\def\Rbar{\vbox{\nis\moveright.20em\vbox{
        \hrule width.50em height.04em}\nis\kern.05em\hbox{$R$}}{}}
\def\Vbar{\vbox{\nis\moveright.05em\vbox{
        \hrule width.60em height.04em}\nis\kern.05em\hbox{$V$}}{}}
\def\Xbar{\vbox{\nis\moveright.20em\vbox{
        \hrule width.60em height.04em}\nis\kern.05em\hbox{$X$}}{}}
\def\thetabar{\vbox{\nis\moveright.15em\vbox{
        \hrule width.30em height.04em}\nis\kern.05em\hbox{$\theta$}}{}}
\def\Lambdabar{\vbox{\nis\moveright.25em\vbox{
        \hrule width.35em height.04em}\nis\kern.05em\hbox{${\mit\Lambda}$}}{}}
\def\Sigmabar{\vbox{\nis\moveright.25em\vbox{
        \hrule width.50em height.04em}\nis\kern.05em\hbox{${\mit\Sigma}$}}{}}
\def\phibar{\vbox{\nis\moveright.18em\vbox{
        \hrule width.40em height.04em}\nis\kern.05em\hbox{$\phi$}}{}}
\def\chibar{\vbox{\nis\moveright.12em\vbox{
        \hrule width.40em height.04em}\nis\kern.05em\hbox{$\chi$}}{}}
\def\psibar{\vbox{\nis\moveright.23em\vbox{
        \hrule width.40em height.04em}\nis\kern.05em\hbox{$\psi$}}{}}
\def\debar{\vbox{\nis\moveright.18em\vbox{
        \hrule width.35em height.04em}\nis\kern.05em\hbox{$\partial$}}{}}
\def\delbar{\vbox{\nis\moveright.10em\vbox{
        \hrule width.63em height.04em}\nis\kern.05em\hbox{$\nabla$}}{}}

\newskip\humongous \humongous=0pt plus 1000pt minus 1000pt

\newif\ifdtup

\begin{document}
\def\thefootnote{\fnsymbol{footnote}}
\begin{center}
{~~~}\\[.2in]
{\large\bf The Fermion Mass Problem}\footnote{Invited talk at the XV
International Warsaw Meeting on Elementary particle Physics, Kazimierz,
Poland, May 25-29, 1992}\\[.3in]
{\bf K.S. Babu}\\[.1in]
{\it Bartol Research Institute}\\[-.07in]
{\it University of Delaware, Newark, DE~~19716}\\[.2in]
{\bf Abstract}\\[.1in]
\end{center}
\baselineskip=12pt
\begin{quotation}

I review the `puzzles' associated with the fermion mass matrices and
describe some recent attempts to resolve them, at least
partially.  Models which attempt to explain the observed mass hierarchy
as arising
from radiative corrections are discussed.  I then scrutinize possible
inter--relations among quark and
lepton masses and the mixing angles
in the context of grand unified theories.
It is argued that the absence of CP violation in the strong
interaction sector (the strong CP problem) may also
have its origin in the structure of the quark mass matrices; such a resolution
does not invoke approximate
global $U(1)$ symmetries resulting in the axion.
Arguments in favor of tiny neutrino masses are
summarized (the solar neutrino puzzle, atmospheric neutrino problem)
and ways to accommodate them naturally are described.

\end{quotation}
\baselineskip=14pt
\noindent{\bf 1. Introduction}

The standard $SU(2)_L \times U(1)_Y$ model of
electro--weak interactions is in remarkably good shape, it has
successfully confronted a wealth of precision data accumulated at the
LEP $e^+e^-$ collider
over the past several years.
The model is beginning to be tested at the
quantum level.  Despite its enormous success and its internal
consistency all the way up to Planckian energies,
it is the widely held view among
theorists that the model will be replaced by a more fundamental theory at
higher energies.  The arguments stem from questions of naturalness in the
symmetry--breaking sector where
the theory is untested.  There is the well--known
fine--tuning problem associated with
fundamental scalars employed
for symmetry breaking, there are other puzzles as well.
The focus of my talk will be the problems associated with the fermion mass
matrices.
There are several facets to the fermion mass puzzle; in a
nutshell, they are our lack of
understanding of (i) family
replication and the resulting proliferation of couplings,
(ii) the observed hierarchy in the fermion masses and mixing angles,
(iii) the origin
of CP violation in weak interactions and its absence in strong
interactions, and (iv) the origin of tiny neutrino masses.
I shall review these problems and describe some recent proposals to
cure one or more of them.

The origin of the mass problem can be traced to family replication,
for with just one generation of fermions
most of the puzzles listed above would not exist.
We know now from the precision measurement of the $Z^0$
width
that there are 3 families with light neutrinos.  All three families
share the same gauge quantum numbers, they are distinguished only by
their Yukawa interactions with the Higgs scalar.  There seems to be no
simple explanation, at least based on the assumption that
quarks and leptons and the Higgs
bosons are elementary, for the multiplicity of generations.  Ideas
based on horizontal symmetries such as $SU(3)_H$ only accommodate
and do not explain
the observed family structure.  Moreover, generating realistic
fermion masses and
mixing angles in such theories is a non--trivial and often complicated
issue.  In what follows, I shall simply accept the existence of three
families and pursue an understanding of the puzzles it creates.

\noindent{\bf 2. Mass and mixing angle hierarchy}

The observed masses of the
three families of quarks and leptons span over five orders of magnitude,
from the electron mass (0.5 MeV) to the top--quark mass
($\ge 91$ GeV).  If neutrinos have masses, as some indirect observations
indicate, then $\nu_e$ should weigh less than 9 eV, which would
further enhance the
hierarchy.  The approximate masses of quarks and leptons
(or bounds in the case of top--quark
and the neutrinos)
are listed in Table 1 where the
light quark masses ($u,d,s)$ are the ones
evaluated at 1 GeV using current algebra.$^1$

\noindent Table 1.  Masses of quarks and leptons in MeV.
\vskip.2in
\begin{center}
\begin{tabular}{|c|c|c|}\hline
$u$ & $c$ & $t$ \\
5.1 & 1270 & $\ge$ 91000 \\
\hline
$d$ & $s$ & $b$ \\
8.9 & 175 & 4250 \\
\hline
$e$ & $\mu$ & $\tau$ \\
0.5 & 106 & 1780 \\
\hline
$\nu_e$ & $\nu_{\mu}$ & $\nu_{\tau}$ \\
$\le 9 \times 10^{-6}$ & $\le$ 0.27 & $\le$ 35 \\
\hline
\end{tabular}
\end{center}
Even if the neutrino masses
are ignored, the Yukawa couplings
that are needed to generate the masses in Table 1 should span from
$h_e \simeq
10^{-6}$ to $h_t \simeq 1$.  The natural value of all couplings is order
$g$, the gauge coupling, which is of order 1.
The puzzle is then why some of the Yukawa's
are as small as $10^{-6}$.

A few features are worth noting regarding the masses in Table 1.
Observe that the charge $2/3$ quark of a
given family is
heavier than the charge $-1/3$ quark, which in turn is heavier than
the corresponding
charged lepton.  This feature is violated only by the first family where
$m_u < m_d$.  Note also that all charged fermions of a given family are
heavier than those of the preceding family--that is, there is no family
overlap.  These features will be helpful in attempts to explain the
hierarchy in a natural manner.

The quark mixing matrix, the Cabibbo-Kobayashi-Maskawa (CKM) matrix, also
exhibits a certain hierarchy, which is likely to be linked to the mass
hierarchy since both the mass eigenvalues and the mixing angles are
obtained from the diagonalization of the same primordial mass
matrices.  The approximate magnitudes of the elements of the
CKM matrix can be displayed in the basis $(u,c,t)^T_L \bullet (d,s,b)_L$ as
\begin{eqnarray}
V=\left(\matrix{0.975 & 0.22 & 0.005 \cr 0.22 & 0.974 & 0.043 \cr
0.01 & 0.04 & 0.99}\right)~~.
\end{eqnarray}
Unitarity of the $3 \times 3$ matrix has been used to write down the
yet to be measured elements of the third row.  Notice that $V$ is
approximately identity, with the leading non--diagonal entry being the
$(12)$ and $(21)$ elements.  The $(23)$ entry is
somewhat smaller and $(13)$ is the smallest.  Indeed, such a hierarchy has
prompted a particularly simple
perturbative parameterization of $V$ due to Wolfenstein:$^2$
\begin{eqnarray}
V = \left(\matrix{1-{1\over 2}{\lambda}^2 & \lambda & A {\lambda}^3
\left(\rho-i \eta\right) \cr
-\lambda & 1-{1 \over 2}{\lambda}^2 & A {\lambda}^2 \cr A {\lambda}^3
\left(1-\rho-i \eta
\right) & -A {\lambda}^2 & 1}\right) + {\cal O}({\lambda}^4)~~,
\end{eqnarray}
where $\lambda\simeq 0.22$ is the Cabibbo angle.  The other parameters
are $A \simeq 1$ and $\sqrt{\rho^2+\eta^2} \simeq 0.5$ with $\eta \ge 0.1$.

In the standard model, both the fermion masses and the quark mixing
angles are free parameters adjusted to their observed
values by hand.  This is clearly  unsatisfactory, since
it leads to a proliferation of parameters.
Although a consistent procedure, it does not explain
why certain Yukawa couplings are orders of magnitude different from others.
A more complete theory, it is hoped, will address these issues and
provide some answers.
Suppose the quark mass matrices have the form prescribed as follows:
\begin{eqnarray}
M_{u,d} = m_{t,b} \left(\matrix{\epsilon^2 & \epsilon^2 & \epsilon^2 \cr
\epsilon^2 & \epsilon & \epsilon \cr \epsilon^2 & \epsilon & 1}\right)
\end{eqnarray}
where $\epsilon \ll 1$.  Numbers of order unity
multiplying various terms with $\epsilon$ in eq. (3) are not displayed.
It is easy to see that the eigen--values of these matrices obey the
hierarchy
\begin{equation}
m_t: m_c: m_u = 1: \epsilon : \epsilon^2~~;~~
m_b: m_s: m_d = 1: \epsilon : \epsilon^2
\end{equation}
The mixing angles have the pattern
\begin{eqnarray}
V_{ud} & \simeq & V_{cs} \simeq V_{tb} \simeq 1 \nonumber \\
V_{us} & \sim & V_{cb} \sim \epsilon ~~;~~
V_{ub} \sim \epsilon^2~~.
\end{eqnarray}
It is apparent that the form of the
matrices in eq. (3) will lead to the
desired hierarchy in the masses as well as in the mixing angles.  The
``average'' value of $\epsilon$
should be around $10^{-2}$ to accommodate both the mixing
and the mass hierarchy.
It is then tempting to postulate that $\epsilon$ is a loop expansion
parameter, $\epsilon \sim (h^2/16 \pi^2)$ with $h$ a typical Yukawa
coupling of order 1.
The third generation masses will then arise at tree--level,
the second family masses arise out of one--loop radiative
corrections and the first family masses are two--loop effects.  There
will be no need to assume any of the couplings to be unnaturally small.
In the
next section, I shall describe a model which illustrates these features.

\noindent{\bf 3. Supersymmetric model for fermion mass hierarchy}

The idea that the second and first family masses arise from
radiative corrections
to some tree--level ``bare''
masses is not new.$^3$  The implementation of this idea has gone through a
revival lately.$^4$
Several models have been proposed in the past five years or so which
attempt to achieve this goal.  In this section, I shall describe one
such attempt.  This is based on work done in collaboration with
B.S. Balakrishna and R.N. Mohapatra.$^5$

The model is based on the supersymmetric
left-right gauge group $SU(2)_L \times SU(2)_R
\times SU(4)_C$.$^6$  There is a reason for
choosing this specific gauge structure, aside from
other well--known motivations for left--right symmetry and
supersymmetry.
The $SU(4)_C$ of color contains a lepto--quark gauge boson
which has an electric charge $2/3$.  Its fermionic superpartner, denoted by
$\lambda$ can mix with the up--type quarks.
This is an attractive feature which can explain why the
top--quark is special, it gets its mass via mixing with the $\lambda$
gaugino.$^7$

The chiral supermultiplets of the model belong to the following
representations ($a=1-3$ is the generation index):
\begin{eqnarray}
\psi_a(2,1,4) = \left(\matrix{u_1 & u_2 & u_3 & \nu \cr
d_1 & d_2 & d_3 & e}\right)_a~~ & ; & ~~ \psi_a^c(1,2,\overline{4}) =
\left(\matrix{u_1^c & u_2^c & u_3^c & \nu^c \cr
d_1^c & d_2^c & d_3^c & e^c}\right)_a  \nonumber \\
N(1,1,6)~~ & ;& ~~ N^c(1,1,6)
\end{eqnarray}
$N$ and $N^c$ each contain a color triplet and a color anti--triplet
fields.  The di--quark (color triplet) and the lepto--quark (color
anti--triplet) components of $N$ will be denoted by $N_1$ and $N_2$.
$N_1^c$ and $N_2^c$ components of $N^c$ have the opposite color
properties.  Under left--right symmetry,
\begin{equation}
\psi_a \leftrightarrow \psi_a^c,~N \leftrightarrow N^c~~.
\end{equation}

Baryon number conservation will be imposed on the model
which prevents $NN$
and $N^cN^c$ terms.
The superpotential of the model has then a very simple form:
\begin{eqnarray}
W &=& {1 \over 4}f_{ab} \left(\psi_a^T i\tau_2 C^{-1}\psi_b N +
\psi_a^{c^T} i \tau_2 \psi_b^c N^c\right) + {1\over 2} M_N N N^c
\nonumber \\
&=& f_{ab}\left(u_ad_bN_1 + u_a e_b N_2-d_a \nu_b N_2 + u_a^c d_b^c N_1^c +
u_a^c e_b^c N_2^c - d_a^c \nu_b^cN_2^c\right) + \nonumber \\
& ~ & M_N\left(N_1N_1^c+N_2N_2^c\right)~~.
\end{eqnarray}
Here in the second line, we have expanded the fields $\psi_a$ and
$\psi_a^c$ in their component
form.  Color indices have been suppressed and summation over the
generation indices
$a,b$ is implied.

Prior to supersymmetry breaking, gauge symmetry remains intact.
Supersymmetry is broken softly with the soft breaking terms given as in
supergravity models.
\begin{equation}
V_{soft} = \sum_{i} \mu_i^2 \phi_i^* \phi_i + m_{3/2} \int d^2\theta
\theta^2 W + \sum_{i=1-3} M_i \lambda_i^T C^{-1} \lambda_i~~.
\end{equation}
Here the sum over $\phi_i$ goes over all the spin zero components of the
chiral superfields.  The second term in eq. (9) stands for the soft SUSY
breaking terms which have the same structure as the superpotential
eq. (8).  The last term is the SUSY--breaking Majorana mass terms for
the gauginos.

$\mu_i^2$ will be chosen negative for $\Tilde{\nu_a}$ and
$\Tilde{\nu_a^c}$ which will lead to spontaneous gauge symmetry
breaking.  The vacuum expectation values (vev's) are denoted by
\begin{equation}
\VEV{\Tilde{\nu^c_a}} = (v_R)_a~;~~ \VEV{\Tilde{\nu_a}} = (v_L)_a~~.
\end{equation}
By performing an orthogonal transformation on the fields $\psi_a$,
the vev's $\VEV{\Tilde{\nu_a}}$ can be brought to the form
\begin{eqnarray}
\left(\matrix{\VEV{\Tilde{\nu_1}} \cr \VEV{\Tilde{\nu_2}} \cr
\VEV{\Tilde{\nu_3}}}\right) = v_L\left(\matrix{0 \cr 0 \cr 1}\right)
= v_L \ket{h}
\end{eqnarray}
A similar rotation on $\psi_a^c$ fields will bring those vev's to
$v_R\ket{h}$.
The quark and lepton masses will be expressed in
terms of the input parameters $f$, which is a $3 \times 3$ Yukawa
coupling matrix, the vector $\ket{h}$ along with $M_N$ and the gaugino
mass.  After symmetry breaking the fields $\Tilde{\nu}$ and
$\Tilde{\nu^c}$ will mix, $\Tilde{u}$ will mix with $\Tilde{u^c}$
(through
$SU(4)_C$ $D$--term); $\Tilde{d}$ mixes with $\Tilde{N_1^c}$ (from the
$F$--term) and with
$\Tilde{N_2}$ (from the soft SUSY breaking term).

Let us turn to the structure of fermion mass matrices in the model.  In
the charge $2/3$ sector, the vev's of $\Tilde{\nu}$ and $\Tilde{\nu^c}$
generate mixing of the three generations with the charge 2/3 gaugino
$\lambda$.  The mixing matrix is given in the basis $(u_a, \lambda)$ as
\begin{eqnarray}
M_{\rm up}=\left(\matrix{0 & g_c\ket{h}v_L \cr \bra{h}v_R g_c & M_{\lambda}}
\right)~~,
\end{eqnarray}
where $g_c$ is the strong gauge coupling.
The matrix of eq. (12) has rank 2.  That is, two of its
eigenvalues are zero.  These two massless states are identified with the
charm and up quarks, the two non--zero masses are given by
\begin{equation}
M^{\prime}_{\lambda} \simeq (M_{\lambda}^2+g_c^2v_R^2)^{1/2}~;~~ m_t \simeq
{{g_c^2 v_L v_R}\over M_{\lambda}^{\prime}}~~~\left(\ket{t} \simeq
\ket{h}\right)~~.
\end{equation}
Note that all three generations were treated on equal footing, yet
only the top--quark acquires a tree--level mass.

In the charge $-1/3$ sector, the mass matrix that mixes the three
generations with the fermions contained in $N,N^c$ is given by
\begin{eqnarray}
M_{\rm down} = \left(\matrix{0 & f \ket{h} v_L \cr \bra{h}f v_R &
M_N}\right)~~.
\end{eqnarray}
Again, this matrix has rank two, meaning that the $d$ and $s$ quark
masses are zero at tree--level.  The $b$ and $N$ masses are given by
\begin{equation}
M_{N_2}^{\prime} = (M_N^2+v_R^2 \VEV{f^2})^{1/2}~;~~ m_b \simeq
\VEV{f^2}
{{v_Lv_R}\over {M_{N_2}^{\prime}}}~~~\left(\ket{b} \simeq f\ket{h}\right)~~.
\end{equation}
Here $\VEV{f^2} \equiv \bra{h}f^2\ket{h}$.
Note that the $b$--quark mass scales inversely with $M_N$ which is a
supersymmetry preserving mass, while $m_t$ scales inversely with the
SUSY--breaking gaugino mass, which is likely to be smaller than $M_N$.
This provides, at least qualitatively, a
reason why top quark is heavier than the bottom.

In the charged lepton sector, since there is no analogous see--saw
partner for the $\tau$ lepton, it remains massless at tree-level.  The
relevant Lagrangian is
\begin{equation}
L_{\rm lepton} = \sqrt{2}gv_Le_3^TC^{-1}\Tilde{W}_L^+ + M_{\Tilde{W_L}}
\Tilde{W}_L^{T^-}C^{-1}\Tilde{W}_L^+ + (L \leftrightarrow R) + h.c.
\end{equation}
Since there is no tree--level mixing between $\Tilde{W}_L$ and
$\Tilde{W}_R$, we concentrate only on the left sector.  It is clear from
eq. (16) that the linear combination defined as
\begin{equation}
\tau = {\rm cos}\theta e_3 -{\rm sin}\theta \Tilde{W}_L^+
\end{equation}
with tan$\theta=\sqrt{2}gv_L/M_{\Tilde{W^+}}$
remains massless, which will be identified with the $\tau$ lepton.

Note that we have not imposed any flavor symmetry on the model.
So nothing will
protect the massless states from acquiring small (and finite) masses
once higher order radiative corrections are included.  The one--loop
diagrams that generate the charm quark and the $\tau$--lepton
masses are shown in fig. 1.  From fig. 1a, one sees
\vskip3.7in
Fig. 1. One--loop graph generating (a) charm mass and
(b), (c) $\tau$ lepton mass.

\noindent that the matrix structure of the one--loop
corrected up--quark matrix is
\begin{equation}
M_{\rm up}^{\rm 1-loop} \sim f\ket{h}\bra{h}f~~.
\end{equation}
This is a unit--rank matrix,
so when added to the rank 2 tree--level matrix, the
rank of the up--quark matrix
increases to three.  That is, the charm quark picks up a mass, but
the up quark still remains massless.
Similarly, both fig. 1b and 1c have the matrix structure $f\ket{h}
\bra{h}f$,
which generates non--zero $\tau$--lepton mass.
$\mu$ and $e$ masses are zero at this stage.
$m_c$ and $m_{\tau}$ can be estimated from fig. 1:
\begin{eqnarray}
m_c & \simeq & {4 \over {16 \pi^2}} \left(\VEV{f^4}-\VEV{f^2}^2\right){{v_Lv_R}
\over {M_N}}~~~\left(\ket{c} \simeq f^2\ket{h} -\VEV{f^2}
\ket{h}\right) \nonumber \\
m_{\tau} & \simeq & {9 \over {16 \pi^2}} \VEV{f^2}g_c^2{{v_Lv_R}\over
{M_N^2}}M_{\lambda}~~~\left(\ket{\tau} \simeq f\ket{h}\right)~~.
\end{eqnarray}
The eigen--states given above are not normalized.

In the down sector, all the one--loop graphs have the matrix structure
given by $f\ket{h}\bra{h}f$, which is the same as the tree-level down quark
matrix.  This implies that one loop corrections do not increase the rank
of the matrix, it only corrects the tree--level $b$--quark mass.
$d$ and $s$ masses remain zero.

Two--loop graphs of fig. 2 will induce non--zero masses for $s$ and $d$
quarks.
\vskip2.5in
Fig. 2. Two--loop diagrams generating (a) strange quark and (b)
$d$--quark mass.

\noindent Fig. 2 has the matrix structure given by
\begin{equation}
M_{\rm down}^{\rm 2-loop} \simeq m_s f^3\ket{h}\bra{h}f^3 + m_d \ket{h}
\bra{h}
\end{equation}
where
\begin{equation}
m_s  \simeq  {{m_c}\over {16 \pi^2}}~;~~
m_d  \simeq  {{\alpha^2}\over{16 \pi^2 {\rm sin}^4\theta_W}}
{{M_{\lambda}^2}\over {M_{W_L}^2M_{W_R}^2}} m_t^2 m_b~~.
\end{equation}
Although $d$ and $s$ quark masses are generated at the same loop--level, there
is a clear distinction between them, each arises from a separate
rank--one graph.
The magnitudes of these graphs are in the right range for acceptable
masses even with all Yukawa couplings being order 1.

In the up--quark sector, there is a diagram analogous to fig. 2b with
$u \leftrightarrow d$ interchange which brings in a new matrix
structure $f\ket{h}\bra{h}f$ in
this sector.  That generates non--zero up--quark mass given by the
second
formula of eq. (21) but with $t$ and $b$ masses interchanged.  That means that
$m_u \propto m_b$ while $m_d \propto m_t$, explaining why
$m_u < m_d$ without further assumption.

In the charged lepton sector, muon and electron masses are
generated at two--loop level
via diagrams analogous to fig 2.  The analog of fig. 2a creates muon mass,
and fig 2b the electron mass.  If the masses of $\Tilde{N_1}$ and
$\tilde{N_2}$ are not split by much, one sees that $m_\mu \simeq m_s$,
which is in good agreement with observations.

Analysis of the neutrino sector at tree--level
involves a $5 \times 5$ matrix
which mixes the gauginos and the neutrinos $\nu_3$ and
$\nu_3^c$.  In the basis $\left(\nu_3,\nu_3^c,\Tilde{W}_{3L},
\Tilde{W}_{3R}, \Tilde{B}\right)$, the matrix reads as
\begin{eqnarray}
\left(\matrix{0 & 0 & gv_L/\sqrt{2} & 0 & -g^{\prime}v_L/\sqrt{2} \cr
0 & 0 & 0 & gv_R/\sqrt{2} & -g^{\prime}v_R/\sqrt{2} \cr
gv_L/\sqrt{2} & 0 & M_L & 0 & 0 \cr
0 & gv_R/\sqrt{2} & 0 & M_R & 0 \cr
-g^{\prime}v_L/\sqrt{2} & -g^{\prime}v_R/\sqrt{2} & 0 & 0 & M_B}\right)
{}~~.
\end{eqnarray}
The lightest eigenstate has a mass given by
\begin{equation}
m_{\nu_3} \simeq {{m_{W_L}^2}\over {M_L}} {\rm tan}\phi~;~~
{\rm tan}\phi = {{g^2M_B+g^{\prime ^2}(M_L+M_R)}\over {g^2M_B+
g^{\prime^2}M_R}}~~.
\end{equation}
Although there is a see--saw type suppression, since the scale $M_L$
cannot be too much higher than a few TeV in this minimal scheme, the
mass of $\nu_{\tau}$ turns out to be in the MeV range unless the angle
$\phi$ is tuned to be small.  Such a neutrino should decay rather fast
in order to avoid cosmological mass density constraints.  One
possibility is that $\nu_3$ decays into
$\rightarrow \nu_{2,1} + \gamma$.

I conclude this section with a few observations.
\begin{enumerate}
\item The idea of generating the mass hierarchy out of radiative
corrections seems to be very promising.  In such a scenario, there is no
need to assume certain Yukawa couplings to be artificially small.  All
couplings can be of the same order and yet
small masses can result due to the
loop suppression factors.
\item It should be emphasized that the mechanism does not rely on any
sort of horizontal symmetry.  All three families of fermions are treated
on par, yet a large mass hierarchy is generated.
\item From phenomenological constraints such as $g-2$ of the $\mu$ and
$\mu \rightarrow e \gamma$ rate, the masses of the scalars should exceed
about 5 TeV.
\item In the example provided above, the weak iso--spin of the scalar
used for mass generation is zero.  This is a desired feature.  If it
were non--zero, flavor-changing couplings of the $Z$ boson with light
quarks will be induced at an unacceptably high level.$^8$
\item An unsatisfactory feature of the idea discussed above is that
if an iso-spin zero scalar is used for generating the charm
quark mass, it becomes proportional to the $b$--quark mass.  Although
there is a log attached to the
loop factor which can be somewhat larger than
unity, since $b$ mass is roughly the same as $c$ mass, $m_c \simeq
\alpha m_b$ is not very desirable.
\item  It has been pointed out$^9$ that in models where there is the
cascade mechanism operative with 1st, 2nd and 3rd family masses induced
at 2, 1 and 0 loop level, the Cabibbo angle tends to be small.  In the
model described above, both $d$ and $s$ quark masses arise at two--loop,
so this problem does not seem to be present.
\end{enumerate}

\noindent{\bf 4. Inter--relations among quark and lepton masses and mixing
angles}

An important aspect of the mass puzzle is
the proliferation of Yukawa couplings.
Perhaps not all parameters are independent, there are sum rules relating
various masses among themselves and with the mixing angles.  The objective
of such an approach is to reduce the number of arbitrary parameters and
make the model more predictive.  In this section I shall review some
popular attempts along this line, especially within the context of grand
unified theories.

In grand unified theories based on simple gauge groups
such as $SU(5)$ or $SO(10)$, quarks and leptons of a given generation
belong to the same grand--unified multiplet.  The
Yukawa couplings of quarks and leptons will then be related.
The best known example is the minimal $SU(5)$ model which predicts
\begin{equation}
M_{\rm down}^0 = M_{\rm lepton}^0~~.
\end{equation}
The superscript $^0$ is to remind ourselves that the equality holds at
the unification scale $\sim 10^{15}~GeV$.   From eq. (24), we have the
asymptotic equality of the eigen--values:
\begin{equation}
m_b^0 = m_\tau^0~;~~ m_s^0 = m_\mu^0~,~~m_d^0 = m_e^0~~.
\end{equation}
To make contact with low
energy observables, we should take into account the evolution of these
mass relations.  If all Yukawa couplings are small,
these relations will evolve linearly which can be integrated
analytically.
However, since we
know that the top Yukawa is non--negligible, the full non--linear
renormalization group equations should be used.  Since measurement of
the weak mixing angle sin$^2 \theta_W$ and the present limits on
proton life--time have excluded minimal $SU(5)$, but are in good
agreement with supersymmetric (SUSY) $SU(5)$, we adhere to the latter.
In the SUSY extension of
the standard model, the evolution of the third family Yukawa couplings
are given by
\begin{eqnarray}
8 \pi^2 {{d h_t^2}\over {dt}} & = & h_t^2\left(6 h_t^2+h_b^2-{13 \over 9}
g_1^2 -3 g_2^2 -{16 \over 3} g_3^2\right) \nonumber \\
8 \pi^2 {{dh_b^2}\over {dt}} & = & h_b^2\left(h_t^2+6 h_b^2+h_\tau^2
-{7 \over 9} g_1^2-3 g_2^2-{16 \over 3} g_3^2\right) \nonumber \\
8 \pi^2 {{d h_\tau^2}\over {dt}} & = & h_\tau^2\left(3 h_b^2 +4 h_\tau^2
-3 g_1^2-3 g_2^2\right)~~.
\end{eqnarray}
These relations can be integrated from a unification scale of $10^{16}$
GeV to the $b$--quark mass scale to arrive at a prediction for
$m_b$.  Results are shown
in Table 2 for varying values of
tan$\beta$  (the ratio of the two Higgs
vev's in the SUSY extension of the standard model) and $m_t$.
The input parameters
used are $\alpha_s(M_Z) = 0.105, \alpha(M_Z) = 1/127.8, \alpha_2(M_Z) =
0.03322$.  The supersymmetric threshold is taken to be at 300 GeV.
These results are to be compared with values of $m_b$ estimated from
quarkonium spectroscopy:$^1$ $m_b(m_b) = 4.25 \pm 0.1~GeV$.  The predictions
for $m_b$ for all values of $m_t$ and tan$\beta$ are in good
agreement with the low energy determination.  Note however that a
heavier top is preferred.

\noindent Table 2.  Predictions for $m_b(m_b)$ from the asymptotic equality
$m_b^0 = m_\tau^0$ in SUSY $SU(5)$ as functions of $m_t$ and tan$\beta$.

$$\begin{array}{|c|c|c|c|c|}\hline
& & & &\\
m_t\rightarrow & 100 & 130 & 160 & 190\\
\hrulefill & & & &\\
\downarrow\!\tan\!\beta & & & &\\\hline
1 & 4.95 & 4.55 & -- & --\\\hline
2 & 5.08 & 4.91 & 4.61 & --\\\hline
3 & 5.08 & 4.95 & 4.72 & 3.78\\\hline
4 & 5.10 & 4.97 & 4.75 & 4.13\\\hline
5 & 5.10 & 4.97 & 4.77 & 4.22\\\hline
6 & 5.10 & 4.97 & 4.77 & 4.26\\\hline
7 & 5.10 & 4.99 & 4.79 & 4.28\\\hline
\end{array}
$$

How about the other mass relations $m^0_s = m_\mu^0$  and $m_d^0=m_e^0$?
 The former would predict
$m_s(1~GeV) \simeq 400 ~MeV$, a factor of 3 larger than the value quoted
in Table 1.  The relation $m_d^0 = m_e^0$ would lead to
$m_d(1~GeV) \simeq 2~MeV$, which is a factor of 3 too small.
This problem can also be seen by noting that
the mass ratios $(m_s/m_d)$
and $(m_\mu/m_e)$ are essentially independent of momentum scale and so
will obey the asymptotic relation which is off by at least a factor of
10.

Modifications of eq. (24) have been proposed to correct the bad mass
relations.  The Georgi--Jarlskog proposal$^{10}$ assumes the mass matrices have
the asymptotic form
\begin{eqnarray}
M_{\rm down} = \left(\matrix{0 & a & 0 \cr a& b & 0 \cr 0 & 0 & c}\right)~;~~
M_{\rm lepton} = \left(\matrix{0 & a & 0 \cr a & -3b & 0 \cr 0 & 0 & c}
\right)~~.
\end{eqnarray}
Such mass matrices can arise if the Higgs sector consists of a {\bf 45}
coupling to the second family
in addition to the usual {\bf 5} and if there are some discrete
symmetries distinguishing generations.  The asymptotic relations implied
by eq. (27) are
\begin{equation}
m_b^0 = m_\tau^0~,;~~ m_s^0 =-{1 \over 3} m_s^0~;~~m_d^0 = -3 m_e^0~~.
\end{equation}
These relations will preserve the successful prediction of $m_b$, in
addition, the $d$ and $s$ quark masses will be predicted to be
$m_d(1~GeV) \simeq 7~ MeV$, $m_s(1~GeV) \simeq 140~MeV$.

It is also possible to predict the quark mixing angles in terms of the
quark mass ratios.  A popular ansatz$^{11}$ that has received much attention
recently
assumes the down quark and charged lepton matrices to have the
Georgi--Jarlskog texture of eq. (27), but the up--quark matrix has the
Fritzsch form:$^{12}$
\begin{eqnarray}
M_{\rm up} = \left(\matrix{0 & a^{\prime} & 0 \cr a^{\prime} & 0 &
b^{\prime} \cr 0 & b^{\prime} & c^{\prime}}\right)~~.
\end{eqnarray}
Although the elements are assumed to be complex, all but one phase can
be rotated away from these matrices.  It can be taken to be the phase of
$a$ in eq. (27).  Since there are only seven parameters describing
thirteen observables, there will be six predictions.  Three of them are
the mass relations of eq. (28), the other three are for the quark mixing
angles.  These asymptotic relations are given by ($^0$ is dropped for
convenience)
\begin{eqnarray}
|V_{us}| = |V_{cd}| = \left|\sqrt{{d \over s}}-\sqrt{{u \over c}}
e^{i\phi}\right|~;~~
|V_{cb}| = |V_{ts}| = \sqrt{{c \over t}} \nonumber \\
|V_{ub}| = \sqrt{{u \over t}}~;~~
|V_{td}| = \left|\sqrt{{u \over t}}{c \over t} + \sqrt{{d \over s}}
\sqrt{{c \over t}}e^{i \phi}\right|~~.
\end{eqnarray}
To see the validity of these relations, one has to extrapolate them to
low energies.  It was noted some time ago$^{13}$ that the quark mixing angles
will run appreciably
with momentum if any of the Yukawa coupling is comparable to
the gauge coupling.  The mass ratios also run.  Fig. 3 exhibits the
behavior of these running for the standard spectrum as well as for the
SUSY spectrum (with tan$\beta=3$).  The Cabibbo angle as well as the
mass ratios involving the first two families do not run.  The running
factors for
$|V_{ub}|,|V_{cb}|,|V_{td}|,|V_{ts}|$ are identical, same is true for
$m_d/m_b$ and $m_s/m_b$ etc.
Note that while in the standard
model, $|V_{cb}|,|V_{ub}|$ increase with energy, the opposite is true
for the SUSY model.  One can infer the low energy prediction for
the mixing angles from fig. 3.  $|V_{cb}|$ turns out to be $\ge 0.052$
and the top mass should be near its fixed point value of 180 GeV.
\vskip3.5in
Fig. 3. Running factors for $|V_{cb}|, |m_s/m_b|, |m_c/m_t|$ defined as
$f(M_X)/f(M_Z)$ for the standard model (SM) and SUSY model with
tan$\beta =3$ (from Ref. 14).

Suppose both the up--quark and the down--quark matrices have the
Fritzsch texture.  Then some of the relations of eq. (30) will be
modified.  Of particular interest is the relation for $|V_{cb}|$ which
now takes the form
\begin{equation}
|V_{cb}| = \left|\sqrt{{s \over b}}-\sqrt{{c \over t}}e^{i \phi}\right|
{}~~.
\end{equation}
Note that the magnitude of the first term is about 0.16, so to get
agreement with observed $|V_{cb}| = 0.043 \pm 0.009$, there should be a
strong cancellation between the two terms.  That sets an upper limit of
about 90 GeV on top mass, which is excluded.  The
renormalization properties of this relation has been studied
recently,$^{14}$
where it was shown that $m_t$ as large as 145 GeV is still admissible in
the SUSY model provided tan$\beta$ is large.

\noindent{\bf 5. The strong CP Problem}

The QCD Lagrangian admits a term
\begin{equation}
L^{\prime}_{\rm QCD} = {{\theta_{\rm QCD}}\over {16 \pi^2}} G \Tilde{G}
\end{equation}
where $G$ is the gluon field strength tensor.  Such a term violates both
Parity and $CP$ symmetries.  However, both symmetries are
broken in the weak interaction sector, so there is no reason why such a
term should not exist.  $\theta_{\rm QCD}$ by itself is
not a physical observable, for a chiral rotation on the quark fields
will change its value.  However, the combination
\begin{equation}
\overline{\theta} = \theta_{\rm QCD} + {\rm Arg}(Det M_q)
\end{equation}
is invariant under such rotations and is observable.  Limits on the
electric dipole moment of the neutron puts a severe
constraint $\overline{\theta}
\le 10^{-9}$.  The strong CP problem is why this parameter, which a
priori is of order one is so small.

Several solutions to the puzzle have been put forth.  The
Peccei-Quinn symmetry, which is an approximate axial symmetry broken
only by non--perturbative instantons, can solve the problem, but at
the expense of introducing a light degree of freedom, the axion.  Here I
wish to discuss another solution to the problem which does not result in
the axion.

$\overline{\theta}$ carries information about the structure of the quark mass
matrix.  In theories which respect $P$ or $CP$, the bare QCD
contribution to $\overline{\theta}$ is zero.  If the determinant of the
quark mass matrix is arranged to be real, the second term will also be zero.
This does not mean that there is no weak CP
violation, it could arise from complex CKM matrix elements.  Finite and
calculable $\overline{\theta}$ will be induced at higher order.  If
these induced $\overline{\theta}$ is less than the present limit, that
would provide a solution to the strong CP problem.$^{15}$  In what follows,
I demonstrate this idea in the context of a Parity invariant
theory.$^{16}$

The model is based on the gauge group $SU(2)_L \times SU(2)_R \times
U(1)$.  The quarks and leptons belong to the left--right symmetric
representations under the gauge group:
\begin{equation}
q_L(2,1,1/6)~;~~q_R(1,2,1/6)~;~~ \psi_L(2,1,-1/2)~;~~ \psi_R(1,2,-1/2)~~
{}.
\end{equation}
In addition, there are these singlet quarks and leptons, one per
generation:
\begin{equation}
P(1,1,2/3)~; ~~N(1,1,-1/3)~;~~E(1,1,-1)~~.
\end{equation}
The Higgs sector of the model is very simple, it consists of
$\chi_L(2,1,1/2)+\chi_R(1,2,1/2)$.  With such a spectrum, the ordinary
quark and lepton masses can arise only via its mixing with the exotic
fermions.$^{17}$  The mass matrix for the up--sector is given by
\begin{eqnarray}
M_{\rm up} = \left(\matrix{0 & hv_L \cr h^{\dagger} v_R & M}\right)
\end{eqnarray}
where $h$ is the $3 \times 3 $ Yukawa coupling matrix.  Although $h$ is
complex, note that the determinant of $M_{\rm up}$ is real, similarly
for $M_{\rm down}$.  So $\overline{\theta}$ is zero at tree level.  Yet
realistic weak CP arises via the KM mechanism.  In this model it turns
out that even at the one--loop level, there is no $\overline{\theta}$
induced.  $\overline{\theta}$ arises at two--loop level, which is
estimated to be $\simeq 10^{-12}$,
well below the present limit.
Although there is no $\overline{\theta}$ at one--loop, the
neutron edm is non--zero, the natural value of it is around $10^{-26}-
10^{-27}~ecm$, which should be accessible to experiments in the near
future.

\noindent{\bf 6. Neutrino masses}

If neutrinos have masses, they are much smaller than their charged
lepton counterparts.  A well--known mechanism that explains this
feature is the see--saw mechanism where the light neutrino mass scales
inversely with the Majorana mass of $\nu_R$.  In $SO(10)$ for
example, they scale as
\begin{equation}
m_{\nu_i} \sim m_{u_i}^2/M~,
\end{equation}
where $M$ is the $\nu_R$ mass, which can be as large as the grand
unification scale.

There are some indirect indications in favor of tiny neutrino masses.
The flux of $\nu_e$'s from the sun detected on earth by several experiments
seems to be smaller than theoretical expectations.  This deficit, known
as the solar neutrino puzzle, can be resolved if neutrinos have tiny
masses and if different flavors mix.  An elegant explanation is the
matter enhanced resonant oscillation (the Mikheyev-Smirnov-Wolfenstein
effect).  The favored value of the neutrino mass splitting and mixing
angle is around $\Delta m^2 \simeq 10^{-5}~
eV^2$ and sin$^22\theta \simeq 10^{-2}$.  Such small values originate
naturally in $SO(10)$.

Another puzzle has emerged in the last several years regarding the flux
of neutrinos from the sky (atmospheric neutrinos) detected by the
Kamiokande and IMB experiments.
There seems to be some discrepancy between the observed and expected
$\mu/e$ flux ratio, which can be interpreted as a deficit of
$\nu_\mu$'s or an excess of $\nu_e$'s.  One
possible explanation which is not inconsistent with the MSW mechanism
for solar neutrinos
is in terms of $\nu_\mu-\nu_\tau$
oscillation.  The relevant mixing angle should then satisfy
sin$^22\theta_{\mu \tau} = (0.42-01)$ which is surprisingly large.  For
comparison, the analog of quark mixing is only $3 \times 10^{-2}$.  The
mass difference should lie in the range $(10^{-3}-0.3)~eV^2$.

Are the two independent observations compatible with predictions of
grand unified theories such as $SO(10)$?  If one tries to explain both
simultaneously, two potential problems arise.  Firstly, as noted above,
a resolution of the atmospheric neutrino puzzle requires rather large
mixing angles.  The mass ratios also are not in the favored range.  Note
that the desired spectrum has $(\nu_e,\nu_\mu,\nu_\tau) \sim (\le
10^{-3}, 10^{-3}, 10^{-1})~eV$.  However, from eq. (37), the naive
expectation is
\begin{equation}
m_{\nu_\mu}/m_{\nu_{\tau}} \sim m_c^2/m_t^2 \sim 10^{-4}
\end{equation}
That is, if solar neutrino puzzle is explained via MSW, $\nu_\tau$ mass
turns out to be in the range of a few eV.

In models with non--minimal Higgs, both features can be accommodated
simultaneously.  Consider the mixing of second and third families alone
for now.  Suppose the mass matrices are given by$^{18}$
\begin{eqnarray}
M_{\rm down} & = & \left(\matrix{D & B \cr B & C}\right)~;~~
M_{\rm up} = \left(\matrix{0 & B^{\prime} \cr B^{\prime} & C^{\prime}}
\right)~;~~
M_{\rm lepton} = \left(\matrix{-3D & -3B \cr -3B & C}\right) \nonumber \\
M_{\nu}^{\rm Dirac} & = & \left(\matrix{0 & -3B^{\prime} \cr -3B^{\prime} &
C^{\prime}}\right)~;~~
M_{\nu}^{\rm Majorana} = \left(\matrix{M_2 & M_3 \cr M_3 & 0}\right)~~.
\end{eqnarray}
Here the elements $C,C^{\prime}$ arise from {\bf 10} of Higgs, whereas
all other entries arise from {\bf 126}.  Such matrices preserve the
successful prediction $m_b^0 \simeq m_\tau^0$.  The see-saw
formula now leads to
\begin{equation}
m_{\nu_\mu}/m_{\nu_\tau} \simeq 81 m_c^2/m_t^2~~.
\end{equation}
Note that the factor of 3 in the off--diagonal element of eq. (39)
appears as a factor of 81 in
eq. (40), which makes the mass ratio come out
right.  Due to the same factor of 3,
the (2-3) mixing also turns out to be large (of order 30 degrees or
more), explaining the atmospheric puzzle.  The solar neutrino deficit
can be explained via $\nu_e-\nu_\mu$ MSW oscillations.

\noindent{\bf 7. Conclusion}

It is necessary to go beyond the standard model to address the puzzles
associated with the fermion mass matrices.   Some of the new ideas that
have emerged recently seem to be very promising.  However,
a fully consistent
picture that explains satisfactorily all the puzzles is still lacking.
More work is needed in these directions.

\noindent{\bf Acknowledgments}

I wish to thank Z. Ajduk and the organizers of the Warsaw meeting for
their warm hospitality at Kazimierz.  It is a pleasure to thank my
collaborators B.S. Balakrishna,
X-G. He, E. Ma, R.N. Mohapatra and Q. Shafi.  This work is
supported in part by Department of Energy Grant \#DE-FG02-91ER406267.

\noindent{\bf References}

\begin{enumerate}
\item J. Gasser and H. Leutwyler, {\it Phys. Rept.} {\bf 87}, 77
(1982).
\item L. Wolfenstein, {\it Phys. Rev. Lett.} {\bf 51}, 1945 (1983).
\item S. Weinberg, {\it Phys. Rev. Lett.} {\bf 29}, 388 (1972); H.
Georgi and S.L. Glashow, {\it Phys. Rev.} {\bf D6}, 2977 (1972) and
{\bf D7}, 2457 (1973); R.N. Mohapatra, {\it Phys. Rev.} {\bf D9}, 3461
(1974); S.M. Barr and A. Zee, {\it Phys. Rev.} {\bf D15}, 2652 (1977)
and {\bf D17}, 1854 (1978); S.M. Barr, {\it Phys. Rev.} {\bf D21}, 1424
(1980) and {\bf D24}, 1895 (1981); R. Barbieri and D.V. Nanopoulos,
{\it Phys. Lett.} {\bf B91}, 1980 and {\bf B95}, 43 (1980); M. Bowick
and P. Ramond, {\it Phys. Lett.} {\bf B103}, 338 (1981).
\item B.S. Balakrishna, {\it Phys. Rev. Lett.} {\bf 60}, 1602 (1988);
B.S. Balakrishna, A.L. Kagan and R.N. Mohapatra, {\it Phys. Lett.}
{\bf B205}, 345 (1988); B.S. Balakrishna and R.N. Mohapatra, {\it Phys.
Lett.} {\bf B216}, 349 (1989); A.L. Kagan, {\it Phys. Rev.} {\bf D40},
173 (1989); K.S. Babu and X-G. He, {\it Phys. Lett.} {\bf B219}, 342
(1989); E. Ma, {\it Phys. Rev. Lett.} {\bf 62}, 1228 (1989); {\bf 63},
1042 (1989) and {\bf 64}, 2866 (1990);
K.S. Babu and E. Ma, {\it Mod. Phys. Lett.} {\bf A4}, 1975
(1989); K.S. Babu and R.N. Mohapatra, {\it Phys. Rev. Lett.} {\bf
64}, 2747 (1990); D. Ng and E. Ma, {\it Phys. Rev.
Lett.} {\bf 65}, 2499 (1990);
S.M. Barr, {\it Phys. Rev. Lett.} {\bf 64}, 353
(1990); R. Rattazzi, {\it Z. Phys.} {\bf C 52}, 575 (1991);
Z. Berezhiani and R. Rattazzi, {\it Phys. Lett.} {\bf B279}, 124
(1992).
\item K.S. Babu, B.S. Balakrishna and R.N. Mohapatra, {\it Phys. Lett.}
{\bf B237}, 221 (1990).
\item J.C. Pati and A. Salam, {\it Phys. Rev.} {\bf D10}, 275 (1974).
\item R. Barbieri and L. Hall, {\it Nucl. Phys.} {\bf B319}, 1 (1989).
\item R. Rattazzi, Ref. 4.
\item H.P. Nilles, M. Olechowski and S. Pokorski, {\it Phys. Lett.}
{\bf B248}, 378 (1990).
\item H. Georgi and C. Jarlskog, {\it Phys. Lett.} {\bf B89}, 297
(1979).
\item S. Dimopoulos, L. Hall and S. Raby, {\it Phys. Rev. Lett.}
{\bf 68}, 1984 (1992); J. Harvey, P. Ramond and D. Reiss, {\it Nucl. Phys.}
{\bf B199}, 223 (1982).
\item H. Fritzsch, {\it Phys. Lett.} {\bf B70}, 436 (1977) and
{\bf 73B}, 317 (1978).
\item K.S. Babu, {\it Z. Phys.} {\bf C 35}, 69 (1987); K. Sasaki,
{\it Z. Phys.} {\bf C 32}, 149 (1986); B. Grzadkowski, M. Lindner and S.
Theisen, {\it Phys. Lett.} {\bf B198}, 64 (1987); M. Olechowski and S.
Pokorski, {\it Phys. Lett.} {\bf B257}, 388 (1991).
\item K.S. Babu and Q. Shafi, Bartol Preprint (1992).
\item A.E. Nelson, {\it Phys. Lett.} {\bf 136B}, 387 (1984); S.M. Barr,
{\it Phys. Rev. Lett.} {\bf 53}, 329 (1984) and {\it Phys. Rev.} {\bf
D30}, 1805 (1984).
\item K.S. Babu and R.N. Mohapatra, {\it Phys. Rev.} {\bf D41}, 1286
(1990).
\item Z. Berezhiani, {\it Phys. Lett.} {\bf 129B}, 99 (1983); D. Chang
and R.N. Mohapatra, {\it Phys. Rev. Lett.} {\bf 58}, 1600 (1987); A.
Davidson and K.C. Wali, {\it Phys. Rev. Lett.} {\bf 59}, 393 (1987); S.
Rajpoot, {\it Phys. Lett.} {\bf B191}, 122 (1987).
\item K.S. Babu and Q. Shafi, {\it Phys. Lett.} {\bf B} (1992) (to be
published).

\end{enumerate}
\end{document}